\documentclass{appolb}
\usepackage{epsfig}

\begin{document}
\title{Photoproduction of the $\eta$ Meson from the Deuteron Near Threshold
\thanks{Presented at the 7th International Workshop on Meson Production,
Properties and Interaction, May 24-28 2002, Krakow, Poland}%
}
\author{Ch.~Elster$^{1,2}$, A.~Sibirtsev$^1$, S.~Schneider$^1$,
J.~Haidenbauer$^1$, S.~Krewald$^1$ and J.~Speth$^1$
\address{
$^1$Institut f\"ur Kernphysik, Forschungszentrum J\"ulich,
D-52425 J\"ulich \\
$^2$Department of Physics and Astronomy,
Ohio University, Athens, OH 45701
}
}
\maketitle
\begin{abstract}
Very recent data for the reaction  $\gamma{d}{\to}\eta{np}$
are analyzed within a model
that includes contributions from the impulse approximation  and
next order corrections due to the $np$ and $\eta{N}$ interactions
in the final state. Comparison between the calculations and
data indicate sizable contributions from the $np$ and $\eta{N}$
final state interactions close to threshold.
\end{abstract}
  
\section{Introduction}
Reactions at threshold are often investigated to learn about the interactions between the
reaction products. The study of meson production close to threshold has several peculiar
features. Only a very limited part of the phase space is available for the reaction
products, hence, only a few partial waves contribute to the observables. However, the small
phase space also yields small cross sections. In the entrance channel one deals with large
momenta, but only with small momenta in the final meson-baryon system.

Indeed, a strong influence of the final state interaction (FSI) on the cross
sections of $\pi$, $\eta$, $\eta^\prime$ and $\omega$-meson production in
nucleon-nucleon ($NN$) collisions was observed in experiments at the IUCF,
COSY and CELSIUS accelerator
facilities~\cite{Meyer,Calen1,Moskal1,Moskal3,Machner}.
With the exception of the $\eta$ channel, those experiments producing mesons
in $NN$ collisions can be described almost perfectly by
theoretical calculations accounting only for the final state
interactions between the nucleons~\cite{Moskal3,Machner}. In case of
$\eta$ production there is evidence that the $\eta N$ FSI could play a role
as well~\cite{Calen1}. 
 
\newpage

\noindent\begin{minipage}[t]{5.5cm}
{\epsfxsize=5.5cm \epsffile{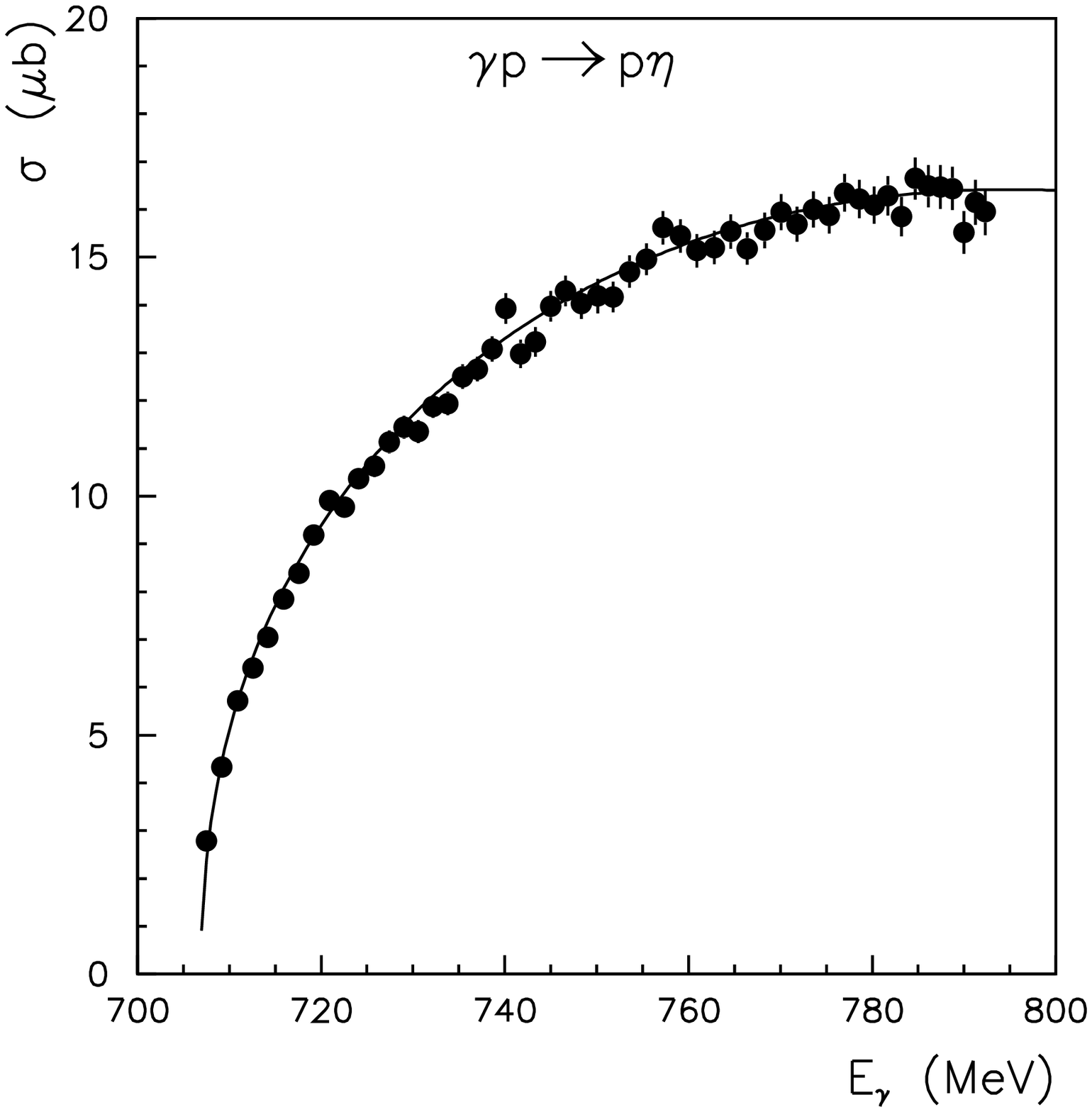}\hspace*{.3cm}}

{\noindent}{{\bf Fig.1:}\footnotesize
 Total cross section for the reaction $\gamma p \rightarrow p \eta$. The experimental
data
are from \protect\cite{Krusche1}, while the solid line gives our result.
}
\end{minipage}

{\vspace*{-7.5cm}\leftskip=6.1cm

Here the reaction $\gamma d \rightarrow np\eta$ \cite{Hejny} 
close to the meson production threshold 
is studied. It offers an opportunity to investigate the final state interactions between
the outgoing particles: proton, neutron, and $\eta$ meson. Provided the FSI between the
nucleons is understood, the reaction allows us to draw conclusions about the $\eta N$
interaction at low energies. In Refs. \cite{eta1,eta2,eta3} we investigated incoherent $\eta$
photoproduction from the deuteron close to threshold taking into account that the reaction
amplitude is given by the sum of the first order term, the impulse approximation (IA), and
the terms of the next higher order due to the final state interactions in the
neutron-proton ($np$) and the $\eta$-nucleon ($\eta N$) system. 

\leftskip=0cm\vspace*{0cm}

As $np$ interaction we
employ one of the high-precision $NN$ potentials, the CD-Bonn potential
\cite{Machleidt1}. The $\eta N$ interaction is extracted from an effective, microscopic
coupled channel model for $\pi N$ scattering developed by the J\"ulich group
\cite{Schutz99,Krehl}. This model includes the  $\eta N$ channel and quantitatively
describes the $\pi N$  phase shifts
and inelasticity parameters in both isospin channels for partial waves up to
$J = {3\over 2}$ and pion-nucleon center-of-mass (c.m.)
energies up to 1.9 GeV \cite{Krehl}. Specifically, it provides a realistic description of
the quantities
relevant for the present investigation, namely  the $S_{11}$ $\pi N$ phase
shift and  the $\pi N \to \eta N$ transition cross section.
Calculations in the same spirit were carried out earlier \cite{Arenhovel1}, however there
the $\eta N$ interaction was treated in a separable ansatz.
In Sec.~2 we  discuss our calculations of the reaction $\gamma d
\rightarrow \eta N$. 
In Sec.~3 we give a phenomenologically based discussion of the enhancement
of $\eta$ meson production in pp collisions in comparison with our results, and we
conclude in Sec.~4.

\section{The Reaction $\gamma d \rightarrow np\eta$}

\subsection{The Elementary Amplitude and the Impuls Approximation}

\noindent
The dominant contribution to the $\eta$-meson photoproduction from a single nucleon
is given by the $N^*$ isobar excitation ~\cite{Knochlein,Benmerrouche}. Since
there is no strong experimental evidence~\cite{PDG3}
for contributions to  $\eta$-meson photoproduction from resonances other than the
$S_{11}(1535)$ isobar in the near-threshold region, we consider only this resonance.
The partial decay widths, $S_{11}(1535){\to}N\eta$ and $S_{11}(1535){\to}N\pi$,
are related to the relevant coupling constant
$g_{RN\xi}$, $\xi{=}\eta,\pi$, by
\begin{equation}
\Gamma_\xi = \frac{g^2_{RN\xi}}{4\pi} \frac{q_\xi (E_N+m_N)}{M_R}.
\end{equation}
Here the momentum $q_\xi$ and the nucleon energy $E_N$ are evaluated in
the rest frame of the resonance at the pole position of $S_{11}(1535)$.
Considering only the contribution of the $S_{11}(1535)$ resonance,
the data for $\eta$-meson photoproduction off protons
can be well fitted with the following resonance parameters
at the $S_{11}(1535)$ pole:  $M_R=1544~MeV, \,\, \Gamma=203~MeV,
\Gamma_\eta/\Gamma=0.45 \,\, \Gamma_\pi/\Gamma=0.45 \,\ ,
\Gamma_{\pi\pi}/\Gamma=0.1$. Further details are given in Ref.~\cite{eta1}.
 The calculation of the cross section for the reaction
$\gamma p \rightarrow p\eta$ is shown in Fig.~1 in comparison with the data.

Using the impulse approximation (IA) the amplitude ${\cal M}_{IA}$ of
the reaction $\gamma{d}{\to}np{\eta}$ for given
spin $S$ and isospin $T$ of the final nucleons
can be written as
\begin{equation}
{\cal M}_{IA}{=}A^T(s_1)\phi(p_2){-}(-1)^{S+T}A^T(s_2)
\phi (p_1).
\end{equation}
Here $\phi(p_i)$ stands for the deuteron wave function, $p_i$ ($i=1,2)$ is the momentum
of the proton or neutron in the deuteron rest frame, and $A^T$ denotes
the isoscalar or isovector $\eta$-meson photoproduction amplitude at
the squared invariant collision energy $s_N$ given by
\begin{equation}
s_N=s-m_N^2-2(E_\gamma+m_d)E_N+2\vec{k}_\gamma\cdot\vec{p}_i.
\end{equation}
The photon momentum is given by ${\vec k}_\gamma$, and 
$s{=}m_d^2{+}2m_dE_\gamma$ stands for the square of the invariant mass.
Details of the
photoproduction amplitude $A^T$ are described in Ref.\cite{eta1}.
The result for the total cross section for the reaction
$\gamma{d}{\to}np\eta$ based on the
impulse approximation is shown as dotted line in
Fig.~2 in comparison to experiment. Though the impulse approximation describes the data
above $\approx 680$~MeV rather well, close to threshold it underestimates them
substantially.

\subsection{The Final State Interactions}

\noindent
The amplitude ${\cal M}_{NN}$ for the $np$ final state
interaction is given by
\begin{equation}
{\cal M}_{NN}=m_N\int dk\, k^2\, \frac{t_{NN}(q,k)\, A^T(s_N^\prime) \,
\phi(p_N^\prime)}{q^2-k^2+i\epsilon},
\label{nnfsi}
\end{equation}
where $q$ is the nucleon momentum in the final $np$ system and
\begin{equation}
\vec{p}_N^{\, \prime}=\vec{k}+\frac{\vec{k}_\gamma -\vec{p}_\eta}{2},
\end{equation}
$p_\eta$ is the $\eta$-meson momentum and
$p_N^\prime{=}|\vec{p}_N^{\,\prime}|$.
The half-shell $np$ scattering matrix $t_{NN}(q,k)$ in the
$^1S_0$ and $^3S_1$ partial waves  was obtained
at corresponding off-shell momenta $k$
from the CD-Bonn potential~\cite{Machleidt1}. 
Finally, the amplitude ${\cal M}_{\eta N}$ for the $\eta$N
final state interaction is given as
\begin{equation}
{\cal M}_{\eta N} {=} \frac{m_N m_\eta}{m_N+m_\eta}\!
\int \! dk\,  k^2  \, \frac{t_{\eta N}(q,k) \,
A^T(s_N^{\prime\prime})\,   \phi (p_N^{\prime\prime})}
{q^2-k^2+i\epsilon},
\label{etanfsi}
\end{equation}
where the $\eta$-meson momenta in the final and intermediate
state of the $\eta{N}$ system are indicated by $q$ and $k$,
$t_{\eta N}(q,k)$ is the half-shell $\eta{N}$  scattering
matrix in the $S_{11}$ partial wave and
\begin{equation}
\vec{p}_N^{\,\prime\prime}= \vec{k} +
\frac{m_N \, (\vec{k}_\gamma-\vec{p}_N)}{m_N+m_\eta},
\end{equation}
where $\vec{p}_N$ is the momentum of the final proton or neutron in the
deuteron rest frame and $m_\eta$ is the $\eta$-meson mass.

\leftskip=-0.1cm 
\begin{minipage}[h]{6.0cm}
\hspace{-0.2cm}{\epsfxsize=6.0cm \epsffile{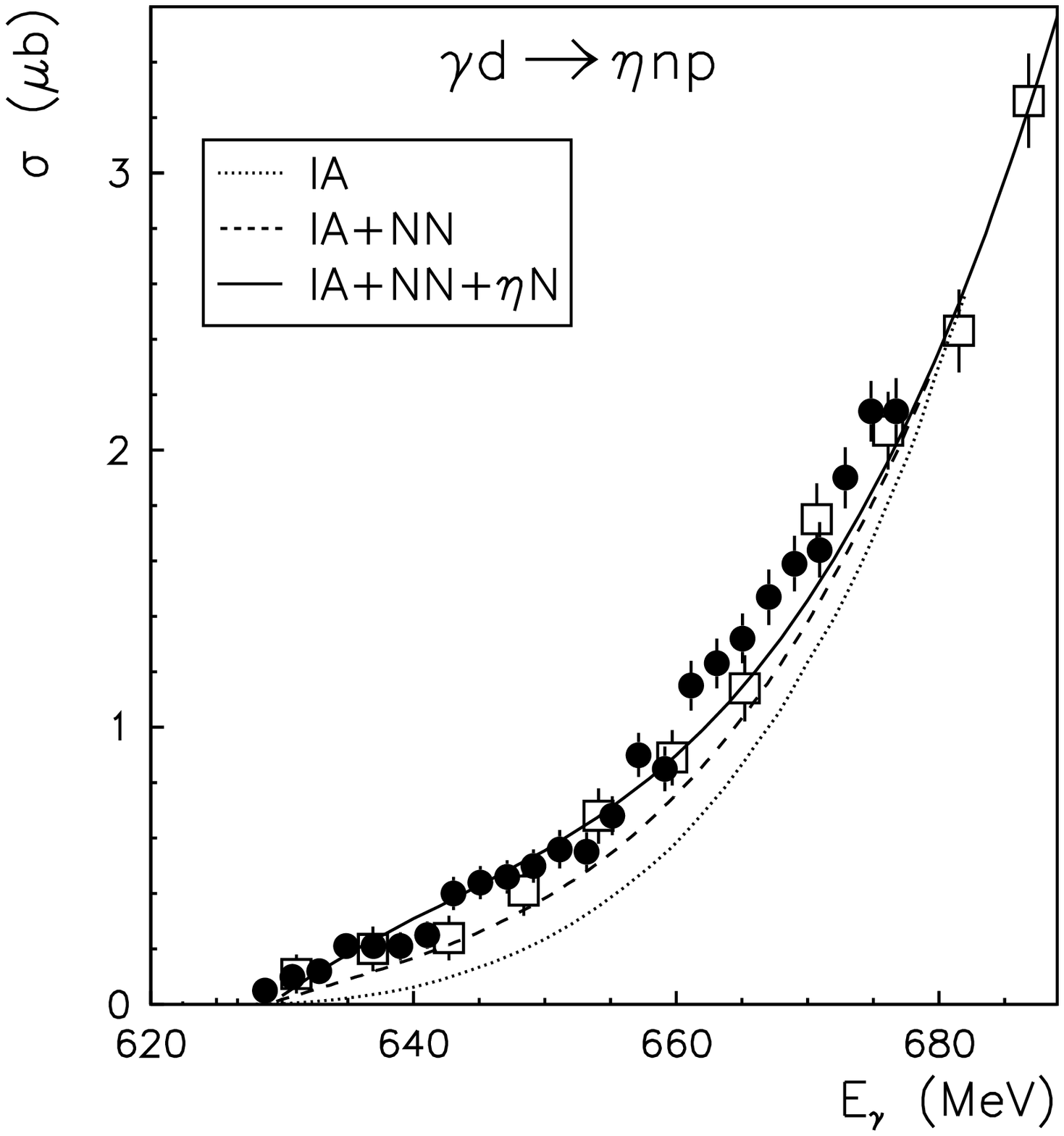}}

{{\bf Fig.2: }\footnotesize
The total cross section for inclusive photoproduction of $\eta$
mesons off deuterium as function of the photon energy $E_{\gamma}$.
The data are from \protect\cite{Hejny} (full circles) and \protect\cite{Krusche1} (open
squares).  The dotted line
represents the IA calculation, while the dashed line is the
result with the $np$ final state interaction. The solid line shows
the full calculation, including the $\eta N$ final state interaction
from the J\"ulich meson-baryon model~\protect\cite{eta2}.
}
\end{minipage}

\vspace*{-11.1cm}\leftskip=6.3cm
\begin{minipage}[h]{6.0cm}
\hspace{-0.3cm}{\epsfxsize=6.0cm \epsffile{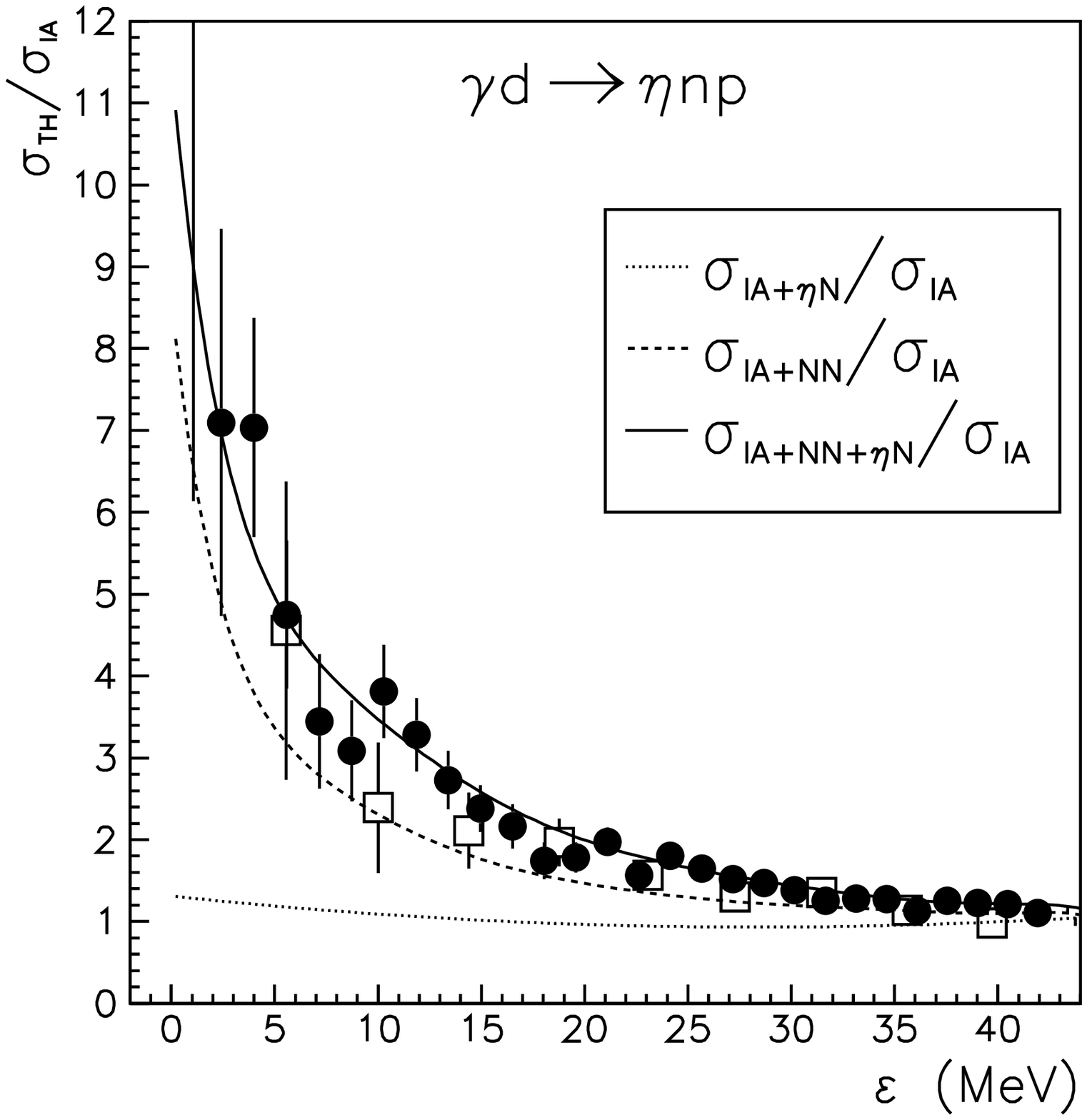}}
{{\bf Fig.3: }\footnotesize
The cross section rations for inclusive photon production of $\eta$-mesons
off the deuteron as a function of the excess energy $\varepsilon = \sqrt{s}-m_p - m_n -
m_\eta $.
Shown are our calculations including the indicated final state interactions
divided by the impulse approximation
The solid line indicates the full calculation containing $np$ and $\eta{N}$
FSI.  The dashed line stands for a calculation including only the $np$ FSI, whereas the
calculation for the dotted line includes only the $\eta{N}$ FSI.
}
\end{minipage}

\leftskip=0cm\vspace*{0.5cm}

The total cross section $\gamma d \rightarrow np\eta$ including the $np$ and 
$\eta N$ FSI in S waves is displayed in Fig.~2.  The dashed line shows the
result for the IA plus $np$ FSI, and the solid line includes in addition the $\eta N$ FSI,
as given by the J\"ulich meson-baryon model. As expected, the attractive $\eta N$ FSI
enhances the cross section very close to the reaction cross section. We observe that for
photon energies larger than 670 MeV there is hardly any effect of the $\eta N$ FSI any
longer. The same is true for the $np$ FSI. Fig.~2 shows that we can well describe the data
close to the reaction threshold, while there is a systematic underprediction of about 10\%
of the experimental results between 660 and 680 MeV photon energy. However, we should not
attribute this discrepancy to the $\eta N$ FSI, since we found in Ref. \cite{eta2} that
the $\eta N$ interaction acts predominantly very close to threshold. We also want to point
out that our calculation matches up with the older data (open squares) at energies larger
than 680 MeV.

It is illuminating to look at the difference in the relative strength of the two
different final state interactions
and their possible interference in our full calculation of the $\eta$
photoproduction cross section. For a more detailed insight, we plot in
Fig.~3 the
ratio of calculations with the final state interactions included separately to the
calculation based on the impulse approximation alone. The dotted line in Fig.~3
represents the calculation including only the $\eta N$ FSI, the dashed line only the
$np$ FSI. From this, it is clear that the $np$ FSI is the dominant one,  a finding already
reported in Ref. \cite{Arenhovel1}. The ratio of the
full calculation containing both, the $\eta N$ and $np$ FSI, to the impulse
approximation
is given by the solid line. A comparison to the two other curves shows that the two
final state interactions interfere constructively at small excess energies,
which magnifies the effect of the relatively weak $\eta N$ FSI.

When characterizing low energy properties of the $\eta N$ interaction within the effective
range approximation, the $\eta N$ on-shell scattering matrix is related to the scattering
length $a_{\eta N}$ as
\begin{equation}
\left[ iq{-}\frac{1}{ a_{\eta N}}\right]^{-1}\!\!\!
{=}\pi \frac{\sqrt{q^2+m_N^2}\sqrt{q^2+m_\eta^2}}
{\sqrt{q^2+m_N^2}+ \sqrt{q^2+m_\eta^2}}\,\, t_{\eta N}(q,q).
\label{efra}
\end{equation}
In previous work \cite{eta2} we showed that within our approach the uncertainty of the
calculations is dominated by the insufficient knowledge of the strength of the $\eta N$
interaction, here represented by $a_{\eta N}$. Moreover, possible
effects due to higher order corrections from the multiple scattering
expansion~\cite{Delborgo,Gillespie} might be overshadowed by the
sizable variation
of $a_{\eta N}$, which as a result of different model calculations or
extractions can range from 0.25+i0.16
to 1.05+i0.27~fm.  The J\"ulich meson-baryon model, whose t-matrix, $t_{\eta N}(q,k)$, we
adopt in our calculations leads to a scattering length $a_{\eta N}$=0.42+i0.32~fm, which
is roughly in the mid-range of the values suggested in the literature.  In Ref. \cite{eta2}
we showed that within our model this value of the scattering length results from the
interplay of resonant, $N^*$(1535), and nonresonant contributions.

We also explored how strongly our calculation of the $\eta$N FSI depends on the specific
properties of the $\eta$N interaction. In order to study whether there is sensitivity
to the off-shell behavior we carried out calculations where the $\eta$N
amplitude of Eq.~(\ref{etanfsi}) is replaced by its effective range expansion of Eq.~(\ref{efra}).
Surprisingly, we obtained  identical results
for the $\eta$ photoproduction cross section
when using either the $\eta$N t-matrix from the effective range
expansion or the one from the full model.
This can be explained through the relatively good representation of
the $\eta{N}$ scattering amplitude by the effective range expansion up to momenta 
$q \approx 350$~MeV, as well as the very weak $k$ dependence of the half-shell $\eta{N}$
t-matrix, $t_{\eta N}(q,k)$ \cite{eta2}.

\begin{minipage}[h]{6.0cm}
\hspace{-0.2cm}{\epsfxsize=6.0cm \epsffile{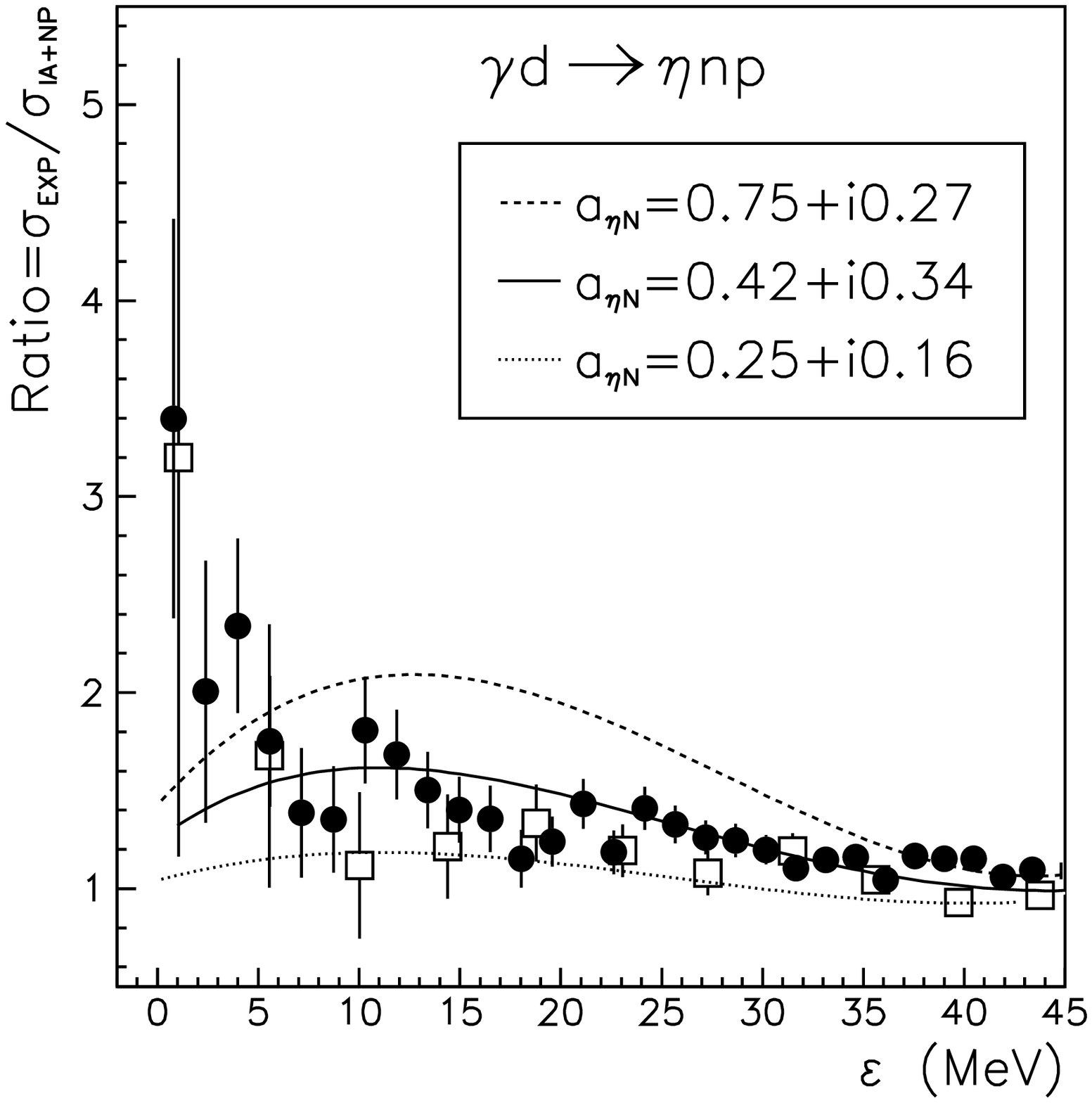}}

{{\bf Fig.4: }\footnotesize
The cross section for the reaction $\gamma d \rightarrow np \eta$ as function of the
excess energy $\varepsilon$. Shown is the experimental cross section divided by our
calculation containing IA and $np$ FSI. The solid line indicates our full calculation
divided by the calculation based on IA and $np$ FSI only. The dotted and dashed lines show
calculations employing the scattering lengths indicated in the figure for the $\eta N$
FSI. The notation for the data is the same as in Fig.~2. 
}
\end{minipage}

\vspace*{-11.1cm}\leftskip=6.3cm
\begin{minipage}[h]{6.0cm}
\hspace{-0.3cm}{\epsfxsize=6.0cm \epsffile{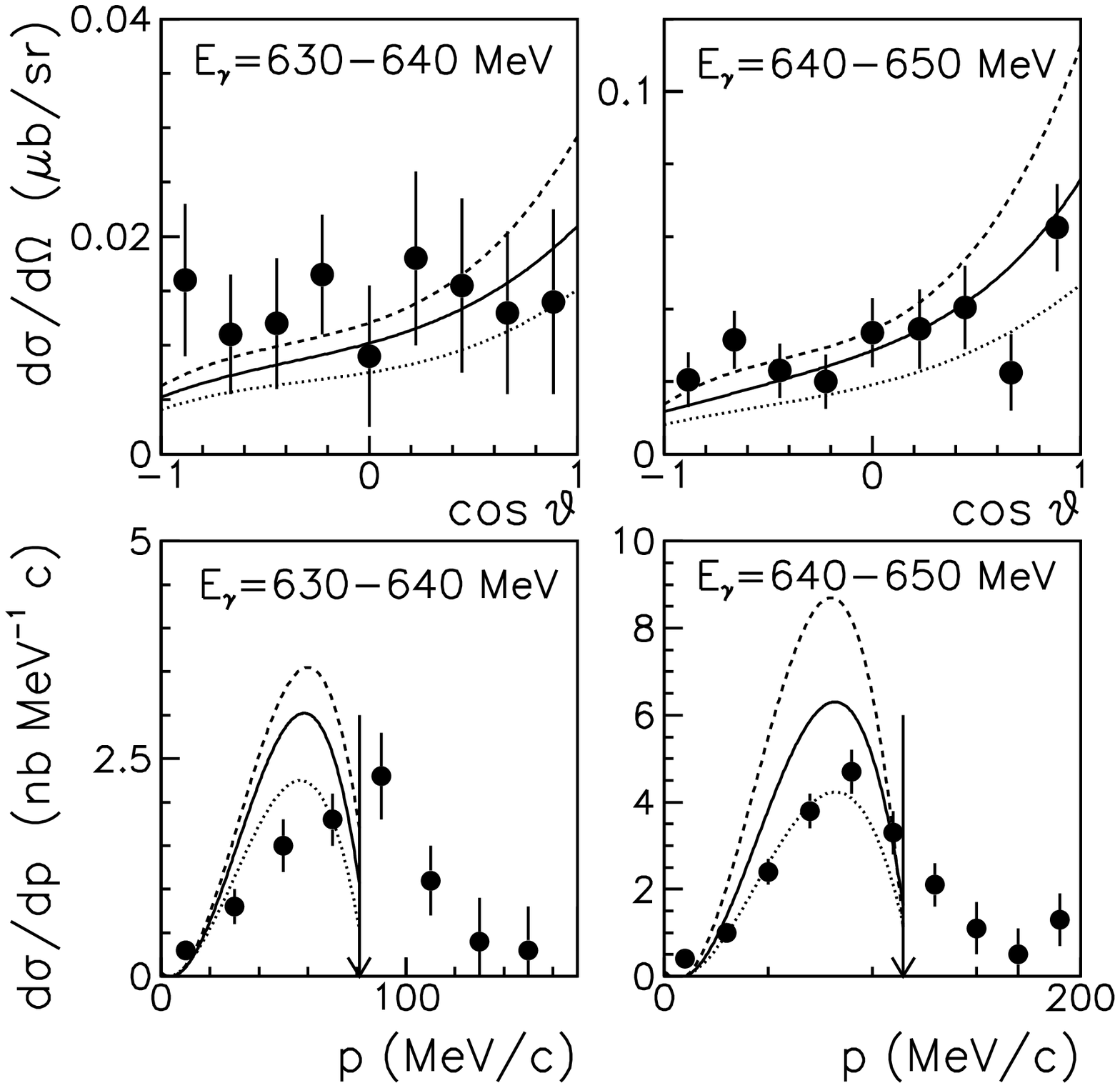}}
{{\bf Fig.5: }\footnotesize
The angular (upper part) and momentum (lower part) spectra of
$\eta$-mesons in the photon-deuteron c.m. system at photon energies
$E_\gamma$=630-640 and 640-650~MeV. The data
are from Ref.~\protect\cite{Hejny}. The lines show our calculations
with different $\eta{N}$ scattering lengths, namely $a_{\eta N}$=0.42+i0.32
(solid), 0.74+i0.27 (dashed) and 0.25+i0.16~fm (dotted). The arrows
indicate the kinematical limit for $\eta$-meson momenta.
}
\end{minipage}

\leftskip=0cm\vspace*{0.5cm}

The next logical step is to see whether some more quantitative information about the
strength of the $\eta{N}$ interaction at low energies can be obtained from the
photoproduction reaction. Since we found that the $\eta{N}$ amplitude obtained with the
effective range expansion is numerically identical to the full calculation of
the amplitude, we can explore the effect of different values for the $\eta{N}$ low energy
parameters on the cross section and momentum distributions close to threshold. 
In Figs. 4 and 5 we display the calculation with our model together with two calculations with
scattering lengths $a_{\eta{N}}{=}0.25{+}i0.16$~fm~\cite{Bennhold} (dotted line) and
$a_{\eta{N}}{=}0.74{+}i0.27$~fm~\cite{Green2} (dashed line). 
The figures suggest that the presently available data for $\varepsilon \leq 40$~MeV show a
preference for smaller values of the $\eta{N}$ scattering length. Similar indications for
a preference for a smaller value of the $a_{\eta N}$ have recently been found in
calculations for the reaction $np \rightarrow d\eta$ close to threshold \cite{pena}.

\section{Phenomenological Consideration of the $\eta N$ FSI in the Reaction $pp \rightarrow
pp\eta$ }

In the previous section we showed that the $\eta N$ final state interaction gives a non
negligible contribution to the cross section of the reaction $\gamma d \rightarrow
np\eta$. Calculations of the reaction $np \rightarrow d\eta$ near threshold also exhibits
considerable sensitivity to the $\eta N$ interaction. One can now ask, if there is some
experimental indication, that the $\eta N$ FSI is also visible close to threshold in the
reaction $pp \rightarrow pp\eta$. 
Recent calculations \cite{nakayama} of that reaction 
include the contribution of the $pp$ FSI but not yet the one of the  $\eta N$ FSI.
Due to the lack of ab initio calculations, one can try if the data themselves reveal
indications of a contribution of the $\eta N$ FSI.

For our estimate of the contribution of the $\eta N$ FSI in the reaction $\gamma d
\rightarrow np \eta$, we divide our full calculation by the calculation that contains
only the IA and the $np$ FSI, see Fig.~4. The deviation from 1 can then be interpreted as
the contribution of the $\eta N$ FSI. In order to proceed in a similar fashion in the
reaction $pp \rightarrow pp\eta$ we first need to extract the contribution of the $pp$
FSI. Since we want to give here phenomenological arguments and do not want to depend on a
specific model, we try to extract it from experiment. For this, we
consider the total cross section of the reaction $pp \rightarrow pp\pi^0$, and
parameterize the $pp$ FSI assuming that the deviation from the phase space is a quadratic
function of the excess energy $\varepsilon$. For small excess energies ($\leq 40$~MeV) this
is certainly justified. The experimental data for the reaction $pp \rightarrow pp\pi^0$
are shown in Fig. 6a. The solid line represents the fit to the deviation from phase space
and is given by
\begin{equation}
f(\varepsilon) = \left( \frac{6}{1+0.042 \varepsilon^2} +1 \right) \cdot \varepsilon^2
\cdot 0.0016 ,
\end{equation}
where the last value is given in $\mu b$. The reaction $pp \rightarrow pp\pi^0$ is chosen,
since it is known that the $\pi N$ FSI is negligible.

Having determined the function $f(\varepsilon)$ representing the $pp$ FSI, we divide
the data for the reaction $pp \rightarrow pp\eta$ by $f(\varepsilon)$ and normalize the
quotient to 1 at an excess energy $\varepsilon \approx 100$~MeV, i.e. far enough away from
the reaction threshold.

\begin{center}
\begin{minipage}[h]{11cm}
\hspace*{-2mm}{\epsfxsize=6.0cm \epsffile{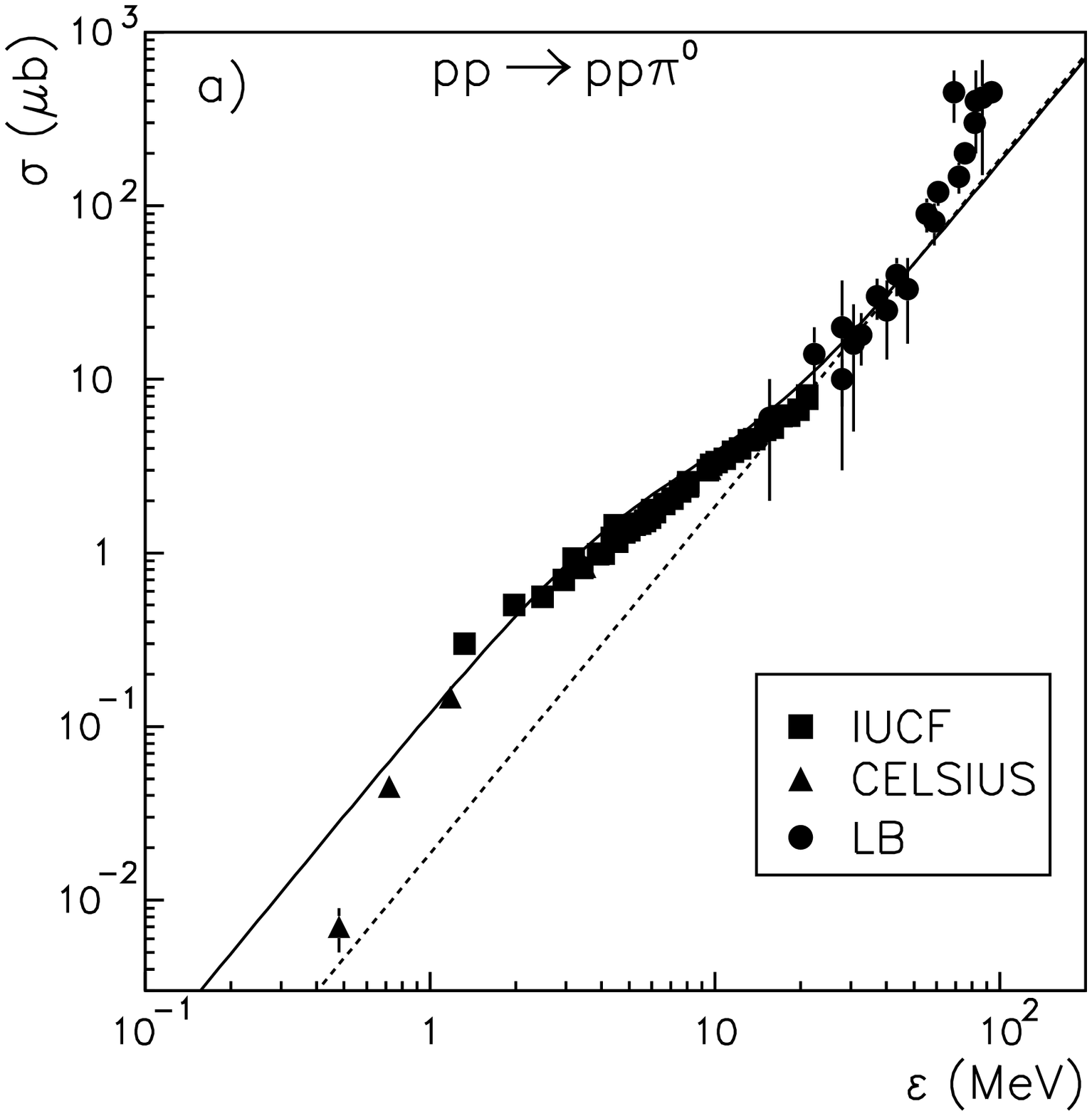}}

\vspace*{-6.1cm}\leftskip=6.1cm
\hspace*{-6mm}{\epsfxsize=6.0cm \epsffile{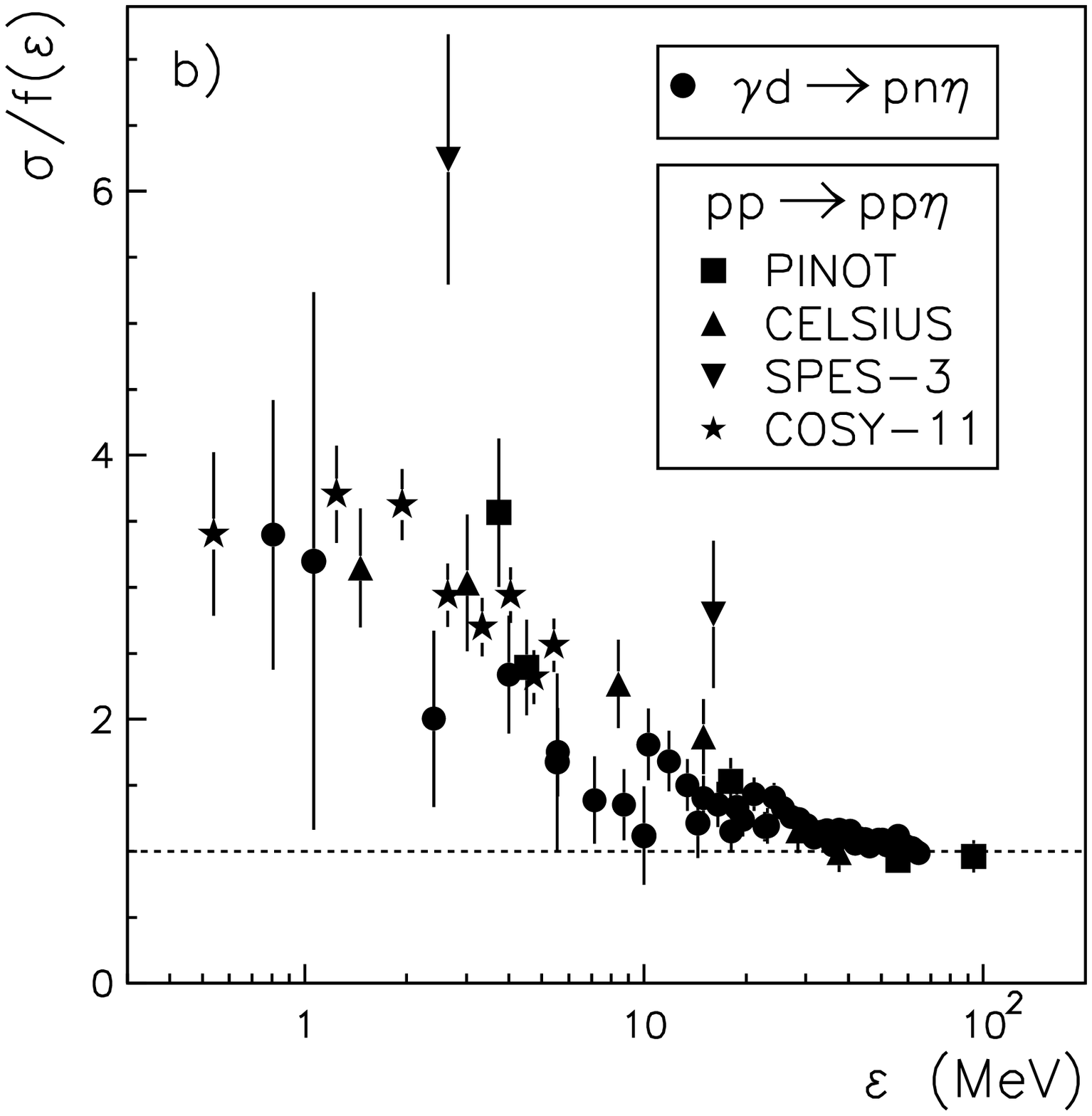}}


\leftskip=0cm\vspace*{0.1cm}

\hspace*{2mm}{{\bf Fig.6: }\footnotesize
The left panel (a) shows the cross section for the reaction $pp \rightarrow pp\pi^0$. 
The solid line gives the function $f(\varepsilon)$ taking into account a deviation from
the phase space proportional to $\varepsilon^2$. The data are from: squares \protect\cite{Meyer},
triangles \protect\cite{bondar}, circles \protect\cite{lb}. 
The right panel (b) shows the total cross section
for the reaction $pp \rightarrow pp\eta$ close to threshold, where the date are divided by
the function $f(\varepsilon)$. Our calculation for the cross section of
 the reaction $\gamma d \rightarrow np\eta$, divided by our calculation based on the IA
and the $np$ FSI alone, are shown by the filled circles. The experimental data are:
squares \protect\cite{pinot}, upward triangles \protect\cite{Calen1}, downward triangles
\protect\cite{spes}, and stars \protect\cite{cosy11}.
}
\end{minipage}
\end{center}

The resulting cross section ratios are depicted in Fig. 6b. For excess energies
$\varepsilon \leq 10$~MeV the data for the reaction $pp \rightarrow pp\eta$ indicate
an enhancement of the cross section. This enhancement is very similar to the one we find
in the reaction $\gamma d \rightarrow np\eta$. For comparison, we plotted those data also
in Fig. 6b, and surprisingly, the enhancement seen in the photoproduction is of the same
order of magnitude as the one phenomenologically extracted from the hadronic reaction.
Of course, the  effect of the $\eta N$ FSI has to be calculated theoretically for the
reaction $pp \rightarrow pp\eta$ in order to draw definite conclusions.

\section{Summary}

We calculated the reaction $\gamma d \rightarrow np\eta$, including the dominant
contribution by the $S_{11}$ resonance in the elementary amplitude and the final state
interactions between all outgoing particles. Those final state interactions influence the
cross section for inclusive photoproduction only for excess energies of the $\eta$ meson
smaller than 40~MeV. At higher energies the cross section is given solely by the impulse
approximation. 

The FSI between the outgoing nucleons is essential to bring the calculated cross section
into the vicinity of the experimental values. Our calculations are based on the CD-Bonn
potential, however due to the presence of the deuteron wave function in the expression of
the FSI amplitude, possible off-shell differences of various $NN$ potential do not
enter the amplitude. Due to a constructive interference effect the $\eta N$ final state
interaction provides an additional enhancement of the production cross section at
energies close to threshold, as required by the data. In our calculations we use the $\eta
N$ interaction extracted from the J\"ulich meson-baryon model. We found that the effect of
the FSI resulting form the $\eta N$ interaction can be very well incorporated into the
model by resorting to an effective range approximation fitted to the full scattering
amplitude of the $\eta N$ model. Guided by this finding, we considered $\eta N$ final state
interactions given by effective range expansions with different values for the scattering
length, and concluded that presently available data for the reaction $\gamma d \rightarrow
np\eta$ are consistent with moderate values of the real part of the scattering length
$a_{\eta N}$. This finding is consistent with a recent calculation of the reaction 
$np \rightarrow d\eta$.


\begin{thebibliography} {99}

\bibitem{Meyer}
        H.O. Meyer et al., Phys. Rev. Lett. \textbf{65}, 2846 (1990);
        H.O. Meyer et al., Nucl. Phys. A \textbf{539}, 633 (1992).
\bibitem{Calen1}
        H. Cal\'en et al., Phys. Lett. B \textbf{366}, 39 (1996);
        Phys. Rev. Lett. \textbf{80}, 2069 (1998).
\bibitem{Moskal1}
        P. Moskal et al., Phys. Rev. Lett. \textbf{80}, 3202 (1998);
        P. Moskal et al., Phys. Lett. B \textbf{474}, 416 (2000).
\bibitem{Moskal3}
        P. Moskal et al., Phys. Lett. B \textbf{482}, 356 (2000).
\bibitem{Machner}
        For an overview and further references see, e.g.,
        H. Machner and J. Haidenbauer, J. Phys. G \textbf{25}, R231 (1999).

\bibitem{Hejny}
        V. Hejny et al., Eur. Phys. J. A{\bf 13}, 493, (2002).
\bibitem{Krusche1}
        B. Krusche et al., Phys. Lett. B \textbf{358}, 40 (1995).

\bibitem{eta1} A. Sibirtsev, Ch. Elster, J. Haidenbauer, J. Speth, Phys. Rev. C{\bf 64},
024006 (2001).
\bibitem{eta2}  A. Sibirtsev, S. Schneider, Ch. Elster, J. Haidenbauer, S. Krewald and J. Speth,
Phys. Rev. C{\bf 65}, 044007 (2002).
\bibitem{eta3} A. Sibirtsev, S. Schneider, Ch. Elster, J. Haidenbauer,
S. Krewald and J. Speth, Phys. Rev. C{\bf 65}, 067002 (2002).

\bibitem{Machleidt1}
        R. Machleidt, Phys. Rev. C \textbf{63}, 024001 (2001).

\bibitem{Schutz99}
        C. Sch\"utz, J. Haidenbauer, J. Speth, and J.W. Durso,
 Phys. Rev. C {\bf 57}, 1464 (1998).
\bibitem{Krehl}
        O. Krehl, C. Hanhart, S. Krewald, J. Speth, Phys. Rev. C {\bf 62},
        025207 (2000).
\bibitem{Arenhovel1}
        A. Fix and H. Arenh\"ovel, Z. Phys. A \textbf{359}, 427 (1997).

\bibitem{Knochlein}
        G. Kn\"ochlein, D. Drechsel and L. Tiator, Z. Phys. A
        \textbf{352}, 327 (1995)
\bibitem{Benmerrouche}
        M. Benmerrouche and N.C. Mukhopadhyay, Phys. Rev. D \textbf{51},
        3237 (1995).
\bibitem{PDG3}
        Particle Data Group, Eur. Phys. J. C \textbf{15}, 1 (2000).

\bibitem{Delborgo}
        R. Delborgo, Nucl. Phys. \textbf{38}, 249 (1962).
\bibitem{Gillespie}
        J. Gillespie, Final State Interactions, Holden-Day (1964) 91.

\bibitem{Bennhold}
        C. Bennhold and H. Tanabe, Nucl. Phys. A {\bf 530}, 625 (1991).
\bibitem{Green2}
        A.M. Green and S. Wycech, Phys. Rev. C {\bf 55}, R2167 (1997).

\bibitem{pena} H. Garcilazo, M.T. Pena, private communication. 

\bibitem{nakayama} K. Nakayama, J. Speth, T.-S.H. Lee, Phys. Rev.  C {\bf 65}, 045210
(2002). 

\bibitem{bondar} A. Bondar {\it et al.}, Phys. Lett {\bf B356}, 8 (1995).
\bibitem{lb} Landolt-B\"ornstein, New Series I/12, edited by H. Schopper, Springer (1988).

\bibitem{pinot} E. Chiavarra {\it et al.}, Phys. Lett {\bf B322}, 270 (1994).
\bibitem{spes} A.M. Bergdolt {\it et al.}, Phys. Rev. D {\bf 48}, 2966 (1993).  
\bibitem{cosy11} J. Smyrski {\it et al.}, Phys. Lett {\bf B474}, 182 (2000).

\end{thebibliography}
\end{document}